\documentclass[preprint]{elsarticle}

\usepackage{amssymb,amsmath}
\usepackage[only,inplus]{stmaryrd}
\usepackage{graphicx}
\newtheorem{thm}{Theorem}
\newtheorem{prp}[thm]{Proposition}
\newdefinition{rmk}{Remark}

\journal{Communications in Nonlinear Science and Numerical Simulation}

\begin{document}

\begin{frontmatter}


\title{Symmetry reduction and exact solutions of the non-linear Black--Scholes equation}

\author[OP1,OP2]{Oleksii Patsiuk\corref{cor1}}
\ead{patsyuck@yahoo.com}
\cortext[cor1]{Corresponding author.}
\author[SK]{Sergii Kovalenko}
\ead{kovalenko@imath.kiev.ua}
\address[OP1]{Institute of Mathematics, National Academy of Sciences of Ukraine, 3 Tereshchenkivs'ka Str., 01601 Kyiv-4, Ukraine}
\address[OP2]{PrivatBank, 6 Seryozhnikova Str., 14006 Chernihiv, Ukraine}
\address[SK]{Ardas Group Inc., 77 20-richchya Peremohy Str., 49127 Dnipro, Ukraine}

\begin{abstract}
In this paper, we investigate the non-linear Black--Scholes equation:
$$u_t+ax^2u_{xx}+bx^3u_{xx}^2+c(xu_x-u)=0,\quad a,b>0,\ c\geq0.$$
and show that the one can be reduced to the equation
$$u_t+(u_{xx}+u_x)^2=0$$
by an appropriate point transformation of variables.
For the resulting equation, we study the group-theoretic properties, namely, we find the maximal algebra of invariance of its in Lie sense, carry out the symmetry reduction and seek for a number of exact group-invariant solutions of the equation. Using the results obtained, we get a number of exact solutions of the Black--Scholes equation under study and apply the ones to resolving several boundary value problems with appropriate from the economic point of view terminal and boundary conditions.
\end{abstract}


\end{frontmatter}


\section{Introduction}\label{1}

In modern mathematical finance, the Black--Scholes equation (BSE) is one of the key equations used in option pricing theory. Note that the standard derivative pricing theory is based on the assumption of perfectly liquid markets. In this case, the well studied linear BSE \cite{BS,M} is used. But in recent years much attention is paid to illiquid markets. As noted in \cite{APS} (see also \cite{BF}), the most comprehensive equation providing the price of a European option is the following non-linear BSE:
\begin{equation}\label{NonLinear}
u_t+\frac12\tilde{\sigma}^2(S,u_S,u_{SS})S^2u_{SS}+r(Su_S-u)=0,\quad r\geq0,
\end{equation}
where $u$ is the price of the European option under study, $S$ is the price of the underlying stock, $r$ is the risk-free interest rate, and $\tilde{\sigma}$ is the volatility function.

For modeling illiquid markets, one can use \cite{BF}:
\begin{enumerate}[1)]
\item transaction-cost models with the volatility function of the form\footnote{This model was suggested by \c Cetin, Jarrow, and Protter \cite{CJP}. Note that in \cite{BF} several other transaction-cost models with some different volatility functions are considered.}:
$$\tilde{\sigma}^2=\sigma^2(1+2\rho Su_{SS});$$
\item reduced-form stochastic differential equation (SDE) models with the volatility function
$$\tilde{\sigma}^2=\frac{\sigma^2}{(1-\rho Su_{SS})^2};$$
\item equilibrium (reaction-function) models with the volatility function
$$\tilde{\sigma}^2=\frac{\sigma^2(1-\rho u_S)^2}{(1-\rho u_S-\rho Su_{SS})^2}.$$
\end{enumerate}
In all these formulas $\sigma$ is the constant (historical) volatility, and $\rho$ is a parameter modeling the liquidity of the market under study\footnote{For $\rho=0$ the market is perfectly liquid (and we have the linear BSE), whereas for $\rho$ large a trade has a substantial impact on the transaction price. For the stock of major U.S. corporations $\rho$ is a small parameter (of the order of $10^{-4}$) \cite[p.~186]{FP}.}.

Since $\rho$ is often considered to be small, we can replace $\tilde{\sigma}^2$ with its first order Taylor approximation around $\rho= 0$ in the last two formulas. Thus, for small values of $\rho$ we can restrict our considerations by the transaction-cost models and investigate only the BSE of the form:
\begin{equation}\label{FinMath}
u_t+\frac12\sigma^2S^2u_{SS}(1+2\rho Su_{SS})+r(Su_S-u)=0,\quad \sigma,\rho>0,\ r\geq0.
\end{equation}

The non-linear BSE with $\tilde{\sigma}^2$ of the form $\sigma^2(1+2\rho Su_{SS})$ is widely used in Financial Mathematics. Note that equation (\ref{FinMath}) is a partial case of equations (1.1) and (28) considered in \cite{AAMR} and \cite{BH}, respectively. Equation (\ref{FinMath}) with $r=0$ was also investigated in \cite{APS, BF, FP, B}. In particular, using methods of the Lie group theory, Bobrov \cite{B} find the maximal algebra of invariance of the one, carry out the symmetry reduction and present examples of exact invariant solutions.

Using the notation $a=\frac12\sigma^2$, $b=\rho\sigma^2$, $c=r$, and $x=S$, we rewrite (\ref{FinMath}) in a more convenient form:
\begin{equation}\label{BSE}
u_t+ax^2u_{xx}+bx^3u_{xx}^2+cxu_x-cu=0,\quad a,b>0,\ c\geq0.
\end{equation}
In what follows, we consider only the values of independent variables $t,x$ from the domain $\mathbb{R}_+\times\mathbb{R}_+$ (this is due to the economic sense of these variables). With a view to avoiding cumbersome calculations made by Bobrov in the case $c=0$, we reduce (\ref{BSE}) to a simpler form using point transformations of variables. Having made the group analysis of the obtained equation and built a number of its exact invariant solutions, we transform them into ones of equation (\ref{BSE}) using the inverse transformations of variables.

The structure of this article is as follows. In Section~\ref{2}, using the simplifying point transformations of variables, we reduce the non-linear BSE~(\ref{BSE}) to a partial differential equation (PDE), which is a special case of an equation from the famous handbook~\cite{PZ}. In Section~\ref{3}, we present the optimal system of the one-dimensional sub-algebras of maximal algebra of invariance (MAI) of the obtained equation, carry out the symmetry reduction, and get a number of exact group-invariant solutions of the one. Returning to BSE~(\ref{BSE}), we obtain a number of its exact solutions in Section~\ref{4}. Next, we apply the solution found in Section~\ref{4} to solving several BVPs with the governing equation~(\ref{BSE}). Finally, in Section~\ref{6}, we briefly sum up the results of this paper.

\section{Simplifying point transformations of variables}\label{2}

Using the point transformations of variables
\begin{gather}
\overline{t}=t,\quad\overline{x}=\log\frac xb,\quad\overline{u}=\frac{bu}x+\frac a2\log\frac xb-\frac{a^2}4t\quad(c=0);\label{trans1}\\
\overline{t}=ct,\quad\overline{x}=\log\frac{cx}b-ct,\quad\overline{u}=\frac{bu}{cx}+\frac a{2c}\log\frac{cx}b-\frac a2\left(1+\frac a{2c}\right)t\quad(c>0),\label{trans2}
\end{gather}
we can reduce equation (\ref{BSE}) to the equation
\begin{equation}\label{ZP}
u_t+(u_x+u_{xx})^2=0
\end{equation}
(hereafter we omit the overlines for convenience).

We get an equation of the form $u_t=F(u_x,u_{xx})$. It is known (see \cite[Subs.~12.1.1,\ No.~2]{PZ}) that the resulting equation admits traveling-wave solution
\begin{equation}\label{wave}
u(t,x)=u(\xi),\quad\xi=kx+\lambda t,
\end{equation}
where the function $u(\xi)$ is determined by the autonomous ordinary differential equation (ODE)
$$F(ku_{\xi},k^2u_{\xi\xi})-\lambda u_{\xi}=0,$$
and a more complicated solution of the form
\begin{equation}\label{complic}
u(t,x)=c_1+c_2t+\varphi(\xi),\quad\xi=kx+\lambda t,
\end{equation}
where the function $u(\xi)$ is determined by the autonomous ODE
$$F(k\varphi_{\xi},k^2\varphi_{\xi\xi})-\lambda\varphi_{\xi}-c_2=0.$$

In Section \ref{3}, we find a number of other solutions of equation (\ref{ZP}).

\section{Symmetry reduction and exact solutions of equation (\ref{ZP})}\label{3}

Using program {\it LIE} \cite{H}, we obtain that the basis of MAI of equation (\ref{ZP}) can be chosen as follows:
$$X_1=-\partial_x,\quad X_2=-e^{-x}\partial_u,\quad X_3=\partial_t,\quad X_4=\partial_u,\quad X_5=t\partial_t-u\partial_u.$$
Non-zero commutators of this operators are:
$$[X_1,X_2]=X_2,\quad[X_2,X_5]=-X_2,\quad[X_3,X_5]=X_3,\quad[X_4,X_5]=-X_4.$$

Hence, our MAI $A$ can be written as a semidirect sum of a one-dimensional algebra and a four-dimensional ideal:
$$A=\{X_5\}\inplus\{X_1,X_2,X_3,X_4\}.$$
The ideal is of the type $A_{2.2}\oplus2A_1$ (here we apply the notations used in \cite{BLZ}). Using this facts and executing the well-known classification algorithm \cite[p.~1450]{PW}, we obtain the following assertion.

\begin{prp}\label{T1}
The optimal system of the one-dimensional subalgebras of MAI of equation (\ref{ZP}) consists of the following ones:
$\langle X_1\rangle$, $\langle X_2\rangle$, $\langle X_3\rangle$, $\langle X_4\rangle$, $\langle X_5\rangle$, $\langle X_1+\varepsilon X_3\rangle$, $\langle X_1+\varepsilon X_4\rangle$, $\langle X_2+\varepsilon X_3\rangle$, $\langle X_2+\varepsilon X_4\rangle$, $\langle X_3+\varepsilon X_4\rangle$, $\langle X_1+y(\varepsilon_1X_3+\varepsilon_2X_4)\rangle$, $\langle X_2+\sin\varphi(\varepsilon_1X_3+\varepsilon_2X_4)\rangle$, $\langle X_5+zX_1\rangle$, $\langle X_5-X_1+\varepsilon X_2\rangle$, where $\varepsilon=\pm1$, $\varepsilon_1=\pm1$, $\varepsilon_2=\pm1$, $y>0$, $z\neq0,-1$, and $0<\varphi<\frac{\pi}2$.
\end{prp}

First of all, note that the algebras $\langle X_2\rangle$, $\langle X_4\rangle$, and $\langle X_2+\varepsilon X_4\ |\ \varepsilon=\pm1\rangle$ do not satisfy the necessary conditions for existence of the non-degenerate invariant solutions. Further, we perform the detailed analysis of invariant solutions, which is based on all other algebras from Proposition \ref{T1}. The results of our investigation are presented in Tables \ref{tab:1} and \ref{tab:2}. Table \ref{tab:1} consists of anzatses generated by the subalgebras and corresponding reduced equations, exact solutions of which (or the first order ODEs, if we could not find their solutions) are given in Table \ref{tab:2}.

\begin{table}
\caption{The symmetry reduction of equation (\ref{ZP})}
\label{tab:1}
\begin{tabular}{llll}
\hline\noalign{\smallskip}
No. & Algebra$^{\rm a}$ & Ansatz & Reduced equation \\
\noalign{\smallskip}\hline\noalign{\smallskip}
1 & $\langle X_1\rangle$ & $u=\varphi(t)$ & $\varphi'=0$ \\
2 & $\langle X_3\rangle$ & $u=\varphi(x)$ & $\varphi''+\varphi'=0$ \\
3 & $\langle X_1+\varepsilon X_3\rangle$ & $u=\varphi(x+\varepsilon t)$ & $(\varphi''+\varphi')^2+\varepsilon\varphi'=0$ \\
4 & $\langle X_1+\varepsilon X_4\rangle$ & $u=\varphi(t)-\varepsilon x$ & $\varphi'=-1$\\
5 & $\langle X_2+\varepsilon X_3\rangle$ & $u=\varphi(x)-\varepsilon te^{-x}$ & $(\varphi''+\varphi')^2-\varepsilon e^{-x}=0$ \\
6 & $\langle X_3+\varepsilon X_4\rangle$ & $u=\varphi(x)+\varepsilon t$ & $(\varphi''+\varphi')^2+\varepsilon=0$\\
7$^{\rm b}$ & $\langle X_1+k(X_3+\varepsilon X_4)\rangle$ & $u=\varphi(y)+\varepsilon t$ & $(\varphi''+\varphi')^2+\frac1k\varphi'+\varepsilon=0$ \\
8$^{\rm c}$ & $\langle X_2+k(X_3+\varepsilon X_4)\rangle$ & $u=\varphi(x)+\left(\varepsilon-\frac1ke^{-x}\right)t$ & $(\varphi''+\varphi')^2-\frac1ke^{-x}+\varepsilon=0$ \\
9 & $\langle X_5\rangle$ & $u=t^{-1}\varphi(x)$ & $(\varphi''+\varphi')^2-\varphi=0$ \\
10$^{\rm d}$ & $\langle X_5+kX_1\rangle$ & $u=t^{-1}\varphi(y)$ & $(\varphi''+\varphi')^2+k\varphi'-\varphi=0$ \\
11$^{\rm e}$ & $\langle X_5-X_1+\varepsilon X_2\rangle$ & $u=e^{-x}(\varphi(y)-\varepsilon x)$ & $(\varphi''-\varphi'+\varepsilon)^2-e^y\varphi'=0$ \\
\noalign{\smallskip}\hline
\multicolumn{4}{l}{$^{\rm a}$\footnotesize In this column, $\varepsilon=\pm1$.}\\
\multicolumn{4}{l}{$^{\rm b}$\footnotesize In this case, $k\neq0$, $y=x+\frac1kt$.}\\
\multicolumn{4}{l}{$^{\rm c}$\footnotesize In this case, $0<|k|<1$.}\\
\multicolumn{4}{l}{$^{\rm d}$\footnotesize In this case, $k\neq0,-1$, $y=x+k\log t$.}\\
\multicolumn{4}{l}{$^{\rm e}$\footnotesize In this case, $y=x-\log t$.}
\end{tabular}
\end{table}

\begin{rmk}
Reduced equations 6 and 8 (with $k<0$) from Table \ref{tab:1} can have real-valued solutions, only if $\varepsilon=-1$.
\end{rmk}

\begin{table}
\caption{The exact group-invariant solutions of equation (\ref{ZP})}
\label{tab:2}
\begin{tabular}{lll}
\hline\noalign{\smallskip}
No. & Algebra$^{\rm a}$ & Exact solution or first order ODE$^{\rm b}$ \\
\noalign{\smallskip}\hline\noalign{\smallskip}
1 & 1 & $u=c_1$ \\
2 & 2 & $u=c_1+c_2e^{-x}$ \\
3 & 3 & $u=c_1-\varepsilon(x+\varepsilon t)+4\delta c_2e^{-\frac12(x+\varepsilon t)}+\varepsilon c_2^2e^{-(x+\varepsilon t)}$ \\
4 & 4 & $u=c_1-t-\varepsilon x$\\
5 & 5 & $u=c_1+4\delta e^{-\frac x2}-(t+c_2)e^{-x}$ \\
6 & 6 ($\varepsilon=-1$) & $u=c_1+c_2e^{-x}+\delta x-t$ \\
7$^{\rm c}$ & 7 ($\varphi'=$ const) & $u=c_1+\varepsilon t+\frac1{2k}\left(\delta\sqrt{1-4\varepsilon k^2}-1\right)\left(x+\frac1kt\right)$ \\
8$^{\rm d}$ & 8 ($\varepsilon=-1$) & $u=c_1+c_2e^{-x}-\left(1+\frac1ke^{-x}\right)t+{}$\\
   & & ${}+\frac{\delta}k\left[\left(k-\frac12e^{-x}\right)x-3\sqrt{k(k+e^{-x})}+
\left(2k-e^{-x}\right)\log\left(\sqrt{k}+\sqrt{k+e^{-x}}\right)\right]$ \\
9$^{\rm e}$ & 8 ($\varepsilon=-1$) & $u=c_1+c_2e^{-x}-\left(1-\frac1ke^{-x}\right)t+{}$ \\
   & & $+\frac{\delta}k\left[\left(k+\frac12e^{-x}\right)x-3\sqrt{k(k-e^{-x})}+\left(2k+e^{-x}\right)\log\left(\sqrt{k}+\sqrt{k-e^{-x}}\right)\right]$ \\
10$^{\rm f}$ & 8 ($\varepsilon=1$) & $u=c_1+c_2e^{-x}+\left(1-\frac1ke^{-x}\right)t+{}$ \\
     & & ${}+\frac{\delta}k\left[2\left(k+\frac12e^{-x}\right)\arctan\sqrt{\frac1k(e^{-x}-k)}-3\sqrt{k(e^{-x}-k)}\right]$ \\
11$^{\rm g}$ & 7 ($\varphi'\neq$ const) & $w'(z)=1-\varepsilon k^2\frac z{w(z)}$ \\
12 & 9 & $w'(z)=\frac1{w(z)}-\sqrt[3]{\frac4{3z}}$ \\
13$^{\rm h}$ & 10 & $w'(\varphi)=\frac{\sqrt{\varphi-kw(\varphi)}}{w(\varphi)}-1$ \\
14 & 11 & $w'(z)=1-\varepsilon\frac{e^{-2z}}{w(z)}$ \\
\noalign{\smallskip}\hline
\multicolumn{3}{l}{$^{\rm a}$\footnotesize In this column, the numbers of algebras from Table \ref{tab:1} are indicated.}\\
\multicolumn{3}{l}{$^{\rm b}$\footnotesize In this column, $\varepsilon,\delta\in\{1,-1\}$; $c_1$, $c_2$ are arbitrary real constants.}\\
\multicolumn{3}{l}{$^{\rm c}$\footnotesize In this case, $k\neq0$, if $\varepsilon=-1$, and $0<|k|\leq\frac12$, if $\varepsilon=1$.}\\
\multicolumn{3}{l}{$^{\rm d}$\footnotesize In this case, $0<k<1$.}\\
\multicolumn{3}{l}{$^{\rm e}$\footnotesize In this case, $0<k<1$, $x\geq-\log k$.}\\
\multicolumn{3}{l}{$^{\rm f}$\footnotesize In this case, $0<k<1$, $x\leq-\log k$.}\\
\multicolumn{3}{l}{$^{\rm g}$\footnotesize In this case, $k\neq0$.}\\
\multicolumn{3}{l}{$^{\rm h}$\footnotesize In this case, $k\neq0,-1$.}
\end{tabular}
\end{table}

\begin{rmk}
In Table \ref{tab:2}:
\begin{enumerate}[1)]
\item solution 1 is trivial and can be included to solution 2;
\item solution 3 is the traveling-wave one, which can be obtained from (\ref{wave}), if we put $k=1$, $\lambda=\varepsilon$;
\item solution 4 can be obtained from solution 3, if we put $c_2=0$;
\item solution 7 is of the form (\ref{complic}), and one can be obtained, if we put $k=1$, $\lambda=\frac1k$, $c_2=\varepsilon$;
\item ODE 11 is obtained, if we put in ODE 7 from Table \ref{tab:1}
$$z=-\frac1ke^{\frac12y},\quad\omega=e^{\frac12y}\sqrt{-\left(\varepsilon+\frac1k\varphi'(y)\right)},$$
and admits the solution in the parametric form (see {\rm\cite[Subs.~1.3.1, No.~2]{ZP}}):
$$z=z(\tau),\quad w=\tau\cdot z(\tau),$$
where $z(\tau)$ is defined as:
\begin{enumerate}[a)]
\item $z(\tau)=c_1\left(\left|2\tau-1+\sqrt{4k^2+1}\right|^{1-\frac1{\sqrt{4k^2+1}}}
\left|2\tau-1-\sqrt{4k^2+1}\right|^{1+\frac1{\sqrt{4k^2+1}}}\right)^{-\frac12}$, if $\varepsilon=-1$, and $k\neq0$;
\item $z(\tau)=c_1\left(\left|2\tau-1+\sqrt{1-4k^2}\right|^{1-\frac1{\sqrt{1-4k^2}}}
\left|2\tau-1-\sqrt{1-4k^2}\right|^{1+\frac1{\sqrt{1-4k^2}}}\right)^{-\frac12}$, if $\varepsilon=1$, and $0<|k|<\frac12$;
\item $z(\tau)=\frac{c_1}{2\tau-1}\,e^{\frac1{2\tau-1}}$, if $\varepsilon=1$, and $k=\pm\frac12$;
\item $z(\tau)=c_1(\tau^2-\tau+k^2)^{-\frac12}\,e^{-\frac1{\sqrt{4k^2-1}}\arctan\frac{2\tau-1}{\sqrt{4k^2-1}}}$, if $\varepsilon=1$, and $|k|>\frac12$;
\end{enumerate}
\item ODE 12 is obtained, if we put in ODE 9 from Table \ref{tab:1}
$$z=\frac16\sqrt{\varphi^3},\quad\omega=\frac12\varphi';$$
this is the Abel equation of the second kind {\rm\cite[Subs.~1.3.2]{ZP}};
\item ODE 13 is obtained, if we put in ODE 10 from Table \ref{tab:1} $w(\varphi)=\varphi'$;
\item ODE 14 is obtained, if we put in ODE 11 from Table \ref{tab:1}
$$z=\frac12y,\quad\omega=e^{-\frac12y}\sqrt{\varphi'(y)}.$$
\end{enumerate}
\end{rmk}

\section{Exact solutions of equation (\ref{BSE})}\label{4}

Using solutions 2--3, 5--10 of equation (\ref{ZP}) (see Table \ref{tab:2}) and the point transformations of variables (\ref{trans1})--(\ref{trans2}), we obtain a number of exact solutions of equation (\ref{BSE}) presented in Tables \ref{tab:3} and \ref{tab:4}.

\begin{table}
\caption{The exact solutions of equation (\ref{BSE}) with $c=0$}
\label{tab:3}
\begin{tabular}{lll}
\hline\noalign{\smallskip}
No. & Sol.$^{\rm a}$ & Exact solution$^{\rm b}$ \\
\noalign{\smallskip}\hline\noalign{\smallskip}
1 & 2 & $u=c_1+\frac a{2b}x\left(c_2+\frac a2t-\log x\right)$ \\
2 & 5 & $u=c_1-t+4\delta\sqrt{\frac xb}+\frac a{2b}x\left(c_2+\frac a2t-\log x\right)$ \\
3 & 6 & $u=c_1+\frac{a+2\varepsilon}{2b}x\left(c_2+\frac{a-2\varepsilon}{2}t-\log x\right)$ \\
4 & 3 & $u=\varepsilon c_1^2e^{-\varepsilon t}+4\delta c_1e^{-\frac{\varepsilon}2t}\sqrt{\frac xb}+
\frac{a+2\varepsilon}{2b}x\left(c_2+\frac{a-2\varepsilon}2t-\log x\right)$ \\
5$^{\rm c}$ & 7 & $u=\frac xb\left[c_1+\left(\varepsilon+\frac{a^2}4\right)t-\frac a2\log x+
\frac1{2k}\left(\delta\sqrt{1-4\varepsilon k^2}-1\right)\left(\frac1kt+\log x\right)\right]$ \\
6$^{\rm d}$ & 8 & $u=c_1-\frac1kt+\frac xb\left\{c_2+\left(\frac{a^2}4-1\right)t-\frac a2\log x+\vphantom{\sqrt{\frac bx}}\right.$ \\
   &    & $\left.{}+\delta\left[\left(1-\frac b{2kx}\right)\log\left(\frac{2kx}b\left(1+\sqrt{1+\frac b{kx}}\right)+1\right)-3\sqrt{1+\frac b{kx}}\right]\right\}$ \\
7$^{\rm e}$ & 9 & $u=c_1+\frac1kt+\frac xb\left\{c_2+\left(\frac{a^2}4-1\right)t-\frac a2\log x+\vphantom{\sqrt{\frac bx}}\right.$ \\
   &    & $\left.{}+\delta\left[\left(1+\frac b{2kx}\right)\log\left(\frac{2kx}b\left(1+\sqrt{1-\frac b{kx}}\right)-1\right)-3\sqrt{1-\frac b{kx}}\right]\right\}$ \\
8$^{\rm f}$ & 10 & $u=c_1-\frac1kt+\frac xb\left\{c_2+\left(\frac{a^2}4+1\right)t-\frac a2\log x+\vphantom{\sqrt{\frac bx}}\right.$ \\
   &      & $\left.{}+\delta\left[2\left(1+\frac b{2kx}\right)\arctan\sqrt{\frac b{kx}-1}-3\sqrt{\frac b{kx}-1}\right]\right\}$ \\
\noalign{\smallskip}\hline
\multicolumn{3}{l}{$^{\rm a}$\footnotesize In this column, the numbers of solutions of equation (\ref{ZP}) from Table \ref{tab:2} are indicated.}\\
\multicolumn{3}{l}{$^{\rm b}$\footnotesize In this column, $\varepsilon,\delta\in\{1,-1\}$; $c_1$, $c_2$ are arbitrary real constants.}\\
\multicolumn{3}{l}{$^{\rm c}$\footnotesize In this case, $k\neq0$, if $\varepsilon=-1$, and $0<|k|\leq\frac12$, if $\varepsilon=1$.}\\
\multicolumn{3}{l}{$^{\rm d}$\footnotesize In this case, $0<k<1$.}\\
\multicolumn{3}{l}{$^{\rm e}$\footnotesize In this case, $0<k<1$, $x\geq\frac bk$.}\\
\multicolumn{3}{l}{$^{\rm f}$\footnotesize In this case, $0<k<1$, $x\leq\frac bk$.}\\
\end{tabular}
\end{table}

\begin{table}
\caption{The exact solutions of equation (\ref{BSE}) with $c\neq0$}
\label{tab:4}
\begin{tabular}{lll}
\hline\noalign{\smallskip}
No. & Sol.$^{\rm a}$ & Exact solution$^{\rm b}$ \\
\noalign{\smallskip}\hline\noalign{\smallskip}
1 & 2 & $u=c_1e^{ct}+\frac a{2b}x\left(c_2+\frac{a+2c}2t-\log x\right)$ \\
2 & 5 & $u=(c_1-ct)e^{ct}+4\delta e^{\frac c2t}\sqrt{\frac{cx}b}+\frac a{2b}x\left(c_2+\frac{a+2c}2t-\log x\right)$ \\
3 & 6 & $u=c_1e^{ct}+\frac{a+2\varepsilon c}{2b}x\left(c_2+\frac{a+2(1-\varepsilon)c}2t-\log x\right)$ \\
4 & 3 & $u=\varepsilon c_1^2e^{(1-\varepsilon)ct}+4\delta c_1e^{\frac c2(1-\varepsilon)t}\sqrt{\frac{cx}b}+
\frac{a+2\varepsilon c}{2b}x\left(c_2+\frac{a+2(1-\varepsilon)c}2t-\log x\right)$ \\
5$^{\rm c}$ & 7 & $u=\frac{cx}b\left\{c_1+\left[\varepsilon c+\frac a2\left(1+\frac a{2c}\right)\right]t-\frac a{2c}\log x+\vphantom{\frac1k}\right.$ \\
   & & $\left.{}+\frac1{2k}\left(\delta\sqrt{1-4\varepsilon k^2}-1\right)\left[\left(\frac1k-1\right)ct+\log x\right]\right\}$ \\
6$^{\rm d}$ & 8 & $u=\left(c_1-\frac ckt\right)e^{ct}+\frac{cx}b\left\{c_2+\left[\frac a2\left(1+\frac a{2c}\right)-c\right]t-\frac a{2c}\log x+\vphantom{\sqrt{\frac bx}}\right.$ \\
   &    & $\left.{}+\delta\left[\left(1-\frac b{2kcx}e^{ct}\right)\log\left(\frac{2kcx}be^{-ct}\left(1+\sqrt{1+\frac b{kcx}e^{ct}}\right)+1\right)-
   3\sqrt{1+\frac b{kcx}e^{ct}}\right]\right\}$ \\
7$^{\rm e}$ & 9 & $u=\left(c_1+\frac ckt\right)e^{ct}+\frac{cx}b\left\{c_2+\left[\frac a2\left(1+\frac a{2c}\right)-c\right]t-
\frac a{2c}\log x+\vphantom{\sqrt{\frac bx}}\right.$ \\
   &    & $\left.{}+\delta\left[\left(1+\frac b{2kcx}e^{ct}\right)\log\left(\frac{2kcx}be^{-ct}\left(1+\sqrt{1-\frac b{kcx}e^{ct}}\right)-1\right)-
   3\sqrt{1-\frac b{kcx}e^{ct}}\right]\right\}$ \\
8$^{\rm f}$ & 10 & $u=\left(c_1-\frac ckt\right)e^{ct}+\frac{cx}b\left\{c_2+\left[\frac a2\left(1+\frac a{2c}\right)+c\right]t-
\frac a{2c}\log x+\vphantom{\sqrt{\frac bx}}\right.$ \\
   &      & $\left.{}+\delta\left[2\left(1+\frac b{2kcx}e^{ct}\right)\arctan\sqrt{\frac b{kcx}e^{ct}-1}-3\sqrt{\frac b{kcx}e^{ct}-1}\right]\right\}$ \\
\noalign{\smallskip}\hline
\multicolumn{3}{l}{$^{\rm a}$\footnotesize In this column, the numbers of solutions of equation (\ref{ZP}) from Table \ref{tab:2} are indicated.}\\
\multicolumn{3}{l}{$^{\rm b}$\footnotesize In this column, $\varepsilon,\delta\in\{1,-1\}$; $c_1$, $c_2$ are arbitrary real constants.}\\
\multicolumn{3}{l}{$^{\rm c}$\footnotesize In this case, $k\neq0$, if $\varepsilon=-1$, and $0<|k|\leq\frac12$, if $\varepsilon=1$.}\\
\multicolumn{3}{l}{$^{\rm d}$\footnotesize In this case, $0<k<1$.}\\
\multicolumn{3}{l}{$^{\rm e}$\footnotesize In this case, $0<k<1$, $x\geq\frac b{kc}e^{ct}$.}\\
\multicolumn{3}{l}{$^{\rm f}$\footnotesize In this case, $0<k<1$, $x\leq\frac b{kc}e^{ct}$.}\\
\end{tabular}
\end{table}


Compare solutions obtained by us with the solutions found in \cite{B}. Changing constants, we can rewrite the Bobrov solutions in the following form:
\begin{gather}
u(t,x)=c_1+c_3x\{c_2+(a-bc_3)t-\log x\};\label{1st}\\
u(t,x)=c_1+c_3t+x\left\{c_2+c_4t-\frac{a}{2b}\log x-\frac{3\delta\sqrt{-c_3bK}}{b}\sqrt{1+\frac1{Kx}}-\right.\notag\\
\left.{}-\frac{\delta c_3K}{\sqrt{-c_3bK}}\left(1-\frac1{2Kx}\right)\log\left[2Kx\left(1+\sqrt{1+\frac1{Kx}}\right)+1\right]\right\},\label{2nd}\\
K=\frac{4c_4b-a^2}{4c_3b},\ \delta=\pm1.\notag
\end{gather}

It is easy to see that solutions 1 and 3 (and also 5 with $a=1$) from Table \ref{tab:3} are of the form (\ref{1st}), and solution 6 is of the form (\ref{2nd}).

\section{Applications to solving various BVPs with the governing PDE (\ref{BSE})}\label{5}

In this section we are going to apply the solutions of the non-linear Black--Sholes equation~\eqref{BSE} found in Section~\ref{4} to solving various BVPs.

In \cite{AAMR} the following stationary BVP for the equation
\begin{equation}\label{5-11}
  \frac{1}{2} \tilde{\sigma}^2 S^2 \dfrac{\partial^2 V}{\partial S^2} + b \sigma^2 S^3 \left( \dfrac{\partial^2 V}{\partial S^2} \right) + r \left( \dfrac{\partial V}{\partial S} S - V \right) = 0, \ \ S \in (c, d)
\end{equation}
under the Dirichlet boundary conditions
\begin{equation}\label{5-12}
  V(c) = V_c, \ \ \ V(d) = V_d
\end{equation}
for some fixed $d > c > 0$ was considered. In equation~\eqref{5-11} the parameter $\tilde{\sigma}^2$ is as follows
\[
  \tilde{\sigma}^2 = \sigma^2 \left( 1 - \dfrac{a}{\sigma} \sqrt{\dfrac{2}{\pi d t}} \right).
\]
The authors proved that under some conditions on the constants $c$, $d$, $V_c$, and $V_d$ the BVP \eqref{5-11} and \eqref{5-12} admits a convex unique classic solution, which can be obtained as the limit of a non-increasing (respectively non-decreasing) sequence of upper (lower) solutions.

Note that in the case $a = 0$, \eqref{5-11} coincides with the stationary version of equation~\eqref{FinMath}. So, it is convenient here to consider the following stationary BVP
\begin{align}
  & a x^2 u_{xx} + b x^3 u_{xx}^2 + c x u_x - cu = 0, \ \ x \in (A, B),\label{5-13}\\
  & \quad x = A: \, u = U_A,\label{5-14}\\
  & \quad x = B: \, u = U_B,\label{5-15}
\end{align}
where $a> 0$, $b > 0$, $c>0$, $B > A \geq 0$, $U_A \geq 0$, and $U_B > 0$. We are going to find an exact solution of the BVP \eqref{5-13}--\eqref{5-15} in explicit form.

Consider solution 5 of Table~\ref{tab:4}:
\begin{multline}\label{5-16}
  u = \dfrac{cx}{b} \left\lbrace  c_1 + \left[ \varepsilon c + \frac{a}{2} \left( 1 + \dfrac{a}{2c} \right) \right] t - \dfrac{a}{2c} \log x \, + \right. \\
  + \left. \dfrac{1}{2k} \left(\delta \sqrt{1 - 4 \varepsilon k^2} - 1 \right) \left[ \left( \frac{1}{k} -1 \right) c t + \log x \right] \right\rbrace, 
\end{multline}
where $c_1 \in \mathbb{R}$, $\varepsilon, \delta \in \{-1, 1\}$, and $k \neq 0$ if $\varepsilon = -1$ or $0 < |k| \leq \frac{1}{2}$ if $\varepsilon = 1$.

This is a solution of the evolution equation~\eqref{BSE}. To obtain a solution of the stationary one~\eqref{5-13}, we need to eliminate the time coefficient in formula~\eqref{5-16}, i.e.
\[
  \varepsilon c + \frac{a}{2} \left( 1 + \dfrac{a}{2c} \right) + \dfrac{c}{2k} \left(\delta \sqrt{1 - 4 \varepsilon k^2} - 1 \right) \left( \frac{1}{k} -1 \right) = 0.
\]
Solving this equation w.r.t. the parameter $k$, we get\footnote{It should be noted that here $k\neq0$, if $\varepsilon=-1$, and $0<|k|\leq\frac12$, if $\varepsilon=1$ (see Table \ref{tab:4}).}
$$k_{1,2}=\frac c2\cdot\frac{\varepsilon c-\frac a2\left(1+\frac a{2c}\right)\pm\left|\varepsilon c+\frac a2\left(1+\frac a{2c}\right)\right|\sqrt{1+\frac{2a}c\left(1+\frac a{2c}\right)}}{(1+\varepsilon)c^2+\varepsilon ac\left(1+\frac a{2c}\right)+\frac{a^2}4\left(1+\frac a{2c}\right)^2}.$$
In this cases, the relevant equation~\eqref{5-13} admits such solution:
\begin{equation}\label{5-17}
  u = \dfrac{c x}{b} \left(c_1 + M \log x \right), 
\end{equation}
where $M \equiv \dfrac{1}{2k} \left(\delta \sqrt{1 - 4 \varepsilon k^2 \mathstrut} - 1 \right) - \dfrac{a}{2c}=\frac{\varepsilon+\frac a{2c}\left(1+\frac a{2c}\right)}{1-\frac1k}-\frac a{2c}$, i.e.,
\begin{equation}\label{5-18}
M_{1,2}=\frac{\varepsilon+\frac a{2c}\left(1+\frac a{2c}\right)}{1-\frac2c\cdot\frac{(1+\varepsilon)c^2+\varepsilon ac\left(1+\frac a{2c}\right)+\frac{a^2}4\left(1+\frac a{2c}\right)^2}{\varepsilon c-\frac a2\left(1+\frac a{2c}\right)\pm\left|\varepsilon c+\frac a2\left(1+\frac a{2c}\right)\right|\sqrt{1+\frac{2a}c\left(1+\frac a{2c}\right)}}}-\frac a{2c}.
\end{equation}

Substitute solution~\eqref{5-17} into the boundary conditions~\eqref{5-14} and \eqref{5-15}, putting in \eqref{5-14} $A=0$ and $U_A=0$. Then \eqref{5-14} is satisfied in the sense of the right limit in $x=0$, and from \eqref{5-15} we obtain the following condition on the coefficient $c_1$:
$$\frac{c B}{b} \left(c_1 + M \log B \right) = U_B.$$
Thus, we get that
\begin{equation}\label{5-19}
c_1 = \frac{bU_B}{cB} - M \log B.
\end{equation}

Hence, we proved the following statement.

\begin{prp}\label{T2}
	The BVP \eqref{5-13}--\eqref{5-15} with $A=0$ and $U_A=0$ admits the classical solution~\eqref{5-17} on $x\in(0,B)$, where the constants $M$ and $c_1$ are defined by formulae~\eqref{5-18} and \eqref{5-19}, respectively.
\end{prp}

The slightly different result is obtained in the case $A > 0$. Here the boundary conditions~\eqref{5-14} and \eqref{5-15} give
\[
  \begin{cases}
    \dfrac{c A}{b} \left(c_1 + M \log A \right) = U_A, \vspace{0.15 cm}\\
    \dfrac{c B}{b} \left(c_1 + M \log B \right) = U_B.
  \end{cases}
\]
From the second condition we still receive formula~\eqref{5-19}, but the first one leads to the condition
\begin{equation}\label{5-20}
  M \log \dfrac{A}{B} + \frac{b}{c} \left( \dfrac{U_B}{B} - \dfrac{U_A}{A} \right) = 0.
\end{equation}

Thus, the following statement is obtained.

\begin{prp}\label{T3}
	The BVP \eqref{5-13}--\eqref{5-15} with $A > 0$ admits the classical solution~\eqref{5-17} on $x\in(A, B)$, where the constants $M$ and $c_1$ are defined by formulae~\eqref{5-18} and \eqref{5-19}, respectively, if and only if condition~\eqref{5-20} holds.
\end{prp}

Now we are going to consider an evolution BVP with the governing equation~\eqref{BSE} on $x \in (0, + \infty)$ and $t \in (0, T)$, the terminal condition
\[
  t = T: \, u = h(x),
\]
and the boundary condition
\[
  x = 0: \, u = 0.
\]

In the terminal condition, $h(x)$ is the so-called pay-off function, which traditionally is taken in the form 
\begin{equation}\label{5-21}
  h(x) = (x - K)^{+},
\end{equation}
and 
\begin{equation}\label{5-22}
h(x) = (K - x)^{+},
\end{equation}
for the European Call and Put options, respectively. Here $K > 0$ is some real constant and the designation
\[ 
 f^{+}(x) =
    \begin{cases}
      f(x)      & \quad \text{if } f(x) > 0, \\
      0         & \quad \text{if } f(x) \leq 0
    \end{cases}
\]
is used.

Note that formulae~\eqref{5-21} and \eqref{5-22} are the simplest forms of the pay-off, which have the strong economic sense, but there are no any evidences against using others, more sophisticated, pay-off functions.

\begin{figure}
\begin{center}
\includegraphics[height=7cm,width=10cm]{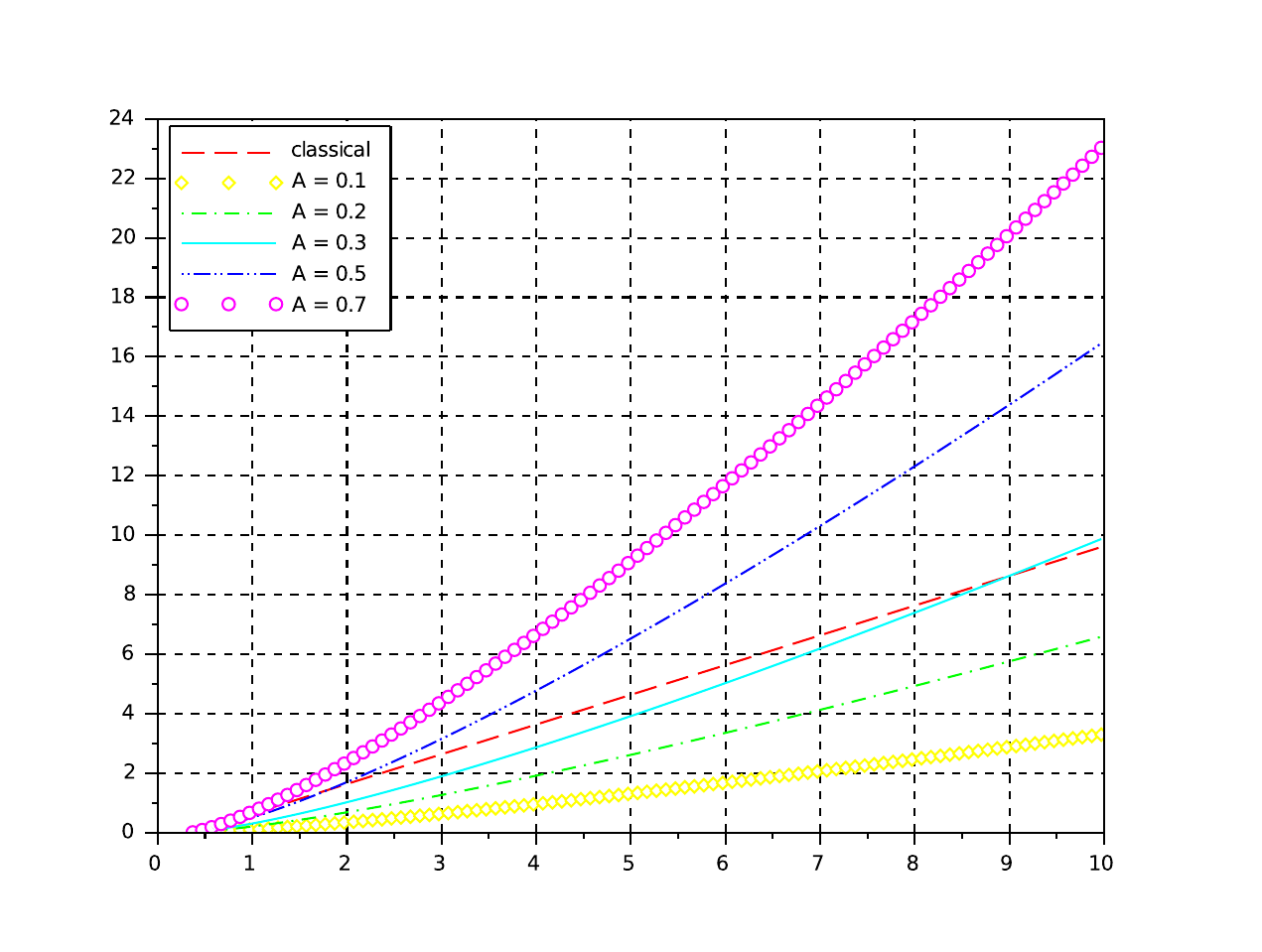}
\caption{Graphs of pay-off functions (\ref{5-21}) for $K=\frac1e$, and (\ref{5-23}) for $B=1$ and various values of $A$}\label{fig:1}
\end{center}
\end{figure}

In our investigation we deals with the following pay-off $h(x)$:
\begin{equation}\label{5-23}
  h(x) = [A x (B + \log x)]^{+},
\end{equation}
where $A > 0$ and $B$ are some real constants. It is easy to see that the behavior of function~\eqref{5-23} is very close to behavior of the classical pay-off function for the European Call option~\eqref{5-21} (see Fig.~\ref{fig:1}).

Thus, we are dealing with such European Call option type BVP
\begin{align}
  & u_t + ax^2u_{xx} + bx^3u_{xx}^2 + cxu_x - cu=0, \ \ x \in (0, + \infty), \ t \in (0, T), \label{5-24}\\
  & \quad t = T: \, u = [A x (B + \log x)]^{+}, \label{5-25}\\
  & \quad x = 0: \, u = 0, \label{5-26}
\end{align}
where $a > 0$, $b > 0$, $c > 0$, $A > 0$, and $B$ are some real constants.

We again intend to use solution 5 of Table~\ref{tab:4} (see formula~\eqref{5-16}). For convenience, we rewrite this solution in the form
\begin{equation}\label{5-27}
  u = \dfrac{c x}{b} \, (C_1 + M (t - T) + N \log x), 
\end{equation} 
where
\[
  M = \varepsilon c + \frac{a}{2} \left( 1 + \dfrac{a}{2c} \right) + \dfrac{c}{2k} \left(\delta \sqrt{1 - 4 \varepsilon k^2} - 1 \right) \left( \frac{1}{k} -1 \right),
\]
\[
  N = \dfrac{1}{2k} \left(\delta \sqrt{1 - 4 \varepsilon k^2 \mathstrut} - 1 \right) - \dfrac{a}{2c},
\]
and 
\[
  C_1 = c_1 + MT.
\]
We also should remind the readers that in \eqref{5-27} $k\neq0$, if $\varepsilon=-1$, and $0<|k|\leq\frac12$, if $\varepsilon=1$.

In view of the terminal condition~\eqref{5-25}, we are looking for a solution of the BVP~\eqref{5-24}--\eqref{5-26} in the form
\begin{equation}\label{5-28}
  u =
    \begin{cases}
      \dfrac{c x}{b} \, (C_1 + M (t - T) + N \log x)     & \quad \text{if } x > e^{-B+\frac MN(T-t)}, \\
      0         & \quad \text{if } 0 < x \leq e^{-B+\frac MN(T-t)}.
    \end{cases}
\end{equation}
Substituting formula~\eqref{5-28} in the terminal condition~\eqref{5-25} and the boundary one~\eqref{5-26}, we find that
\begin{equation}\label{5-29}
  k = - \dfrac{2 c \, (2 b A + a)}{(2 b A + a)^2 + 4 \varepsilon c^2},
\end{equation}
and
\begin{equation}\label{5-30}
  C_1 = BN.
\end{equation}

Thus, we proved the statement.

\begin{prp}\label{T4}
	The European Call option type BVP~\eqref{5-24}--\eqref{5-26} admits the classical solution
	\[
	  u =
	    \begin{cases}
	      \dfrac{c x}{b} \, (B N + M (t - T) + N \log x)     & \quad \text{if } x > e^{-B+\frac MN(T-t)}, \\
  	      0         & \quad \text{if } 0 < x \leq e^{-B+\frac MN(T-t)},
	    \end{cases}
	\]
	where 
	\[
	  M = \varepsilon c + \frac{a}{2} \left( 1 + \dfrac{a}{2c} \right) + \dfrac{c}{2k} \left(\delta \sqrt{1 - 4 \varepsilon k^2} - 1 \right) \left( \frac{1}{k} -1 \right),
	\]
	\[
	  N = \dfrac{1}{2k} \left(\delta \sqrt{1 - 4 \varepsilon k^2 \mathstrut} - 1 \right) - \dfrac{a}{2c},
	\]
	and $k$ is defined by formula~\eqref{5-29} obeying the additional condition: $k\neq0$, if $\varepsilon=-1$, and $0<|k|\leq\frac12$, if $\varepsilon=1$.
\end{prp}

Note that 
\[
  u \sim x \log x \quad \text{as } x \rightarrow + \infty. 
\]

{\bf Example.} Let $a=2\cdot10^{-2},b=4\cdot10^{-6},c=10^{-1},\delta=1,\varepsilon=-1$. If we put $A=10^4,B=1$, and $T=1$ (one year)\footnote{Note that our parameters $a,b,c$, and $T$ are similar to the ones in \cite[p.~809]{AE}.}, then $k=\frac23,M=-0.064,N=0.4$, and 
\[u(t,x)=
\begin{cases}
400x(25\log x-4t+29)     & \quad \text{if } x > e^{0.16t-1.16}, \\
0         & \quad \text{if } 0 < x \leq e^{0.16t-1.16}.
\end{cases}
\]

\begin{figure}
\begin{center}
\includegraphics[height=7cm,width=10cm]{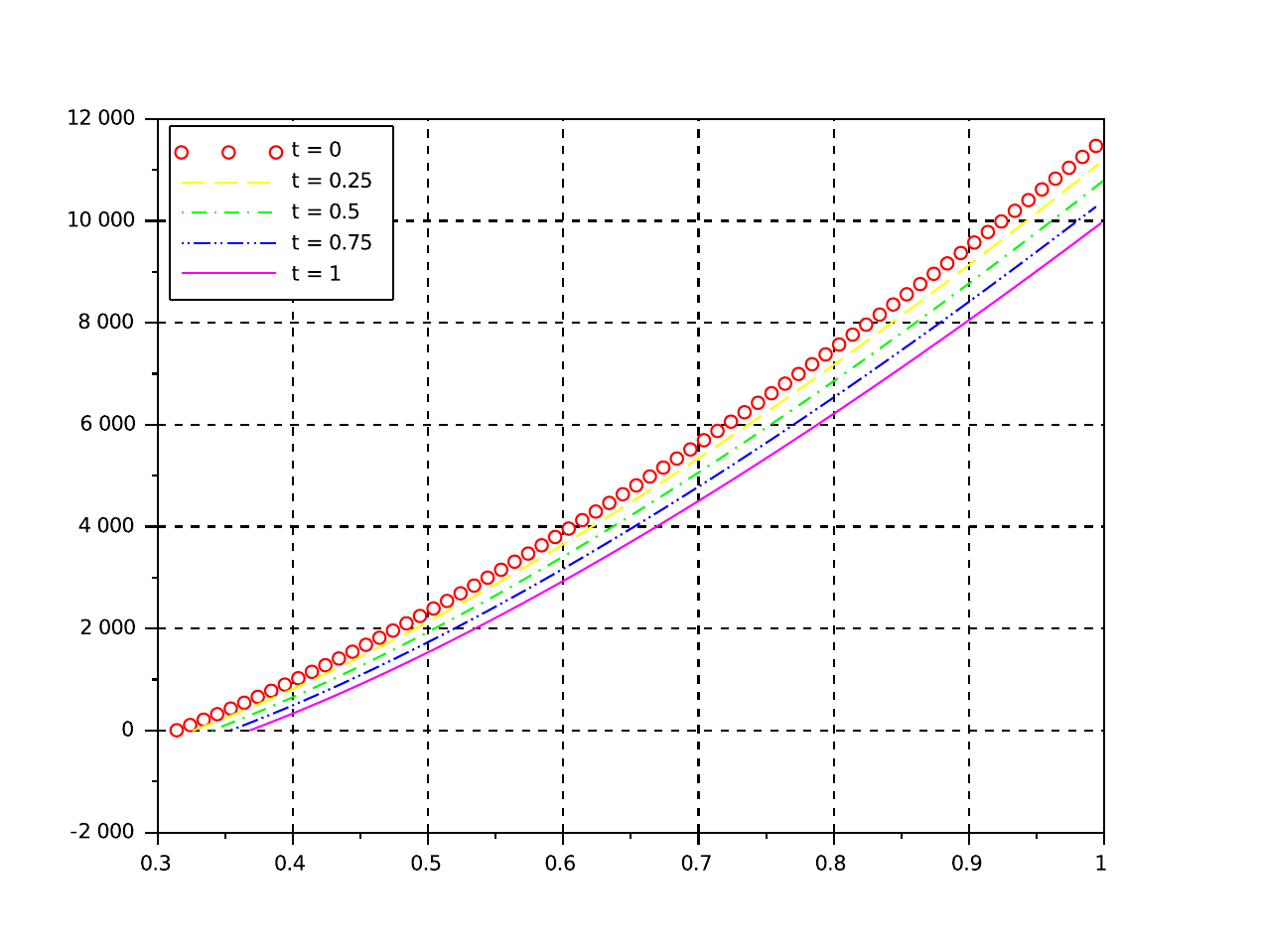}
\caption{Solution of the BVP (\ref{5-24})--(\ref{5-26}) $u=u(t_i,x)$, for some fixed values $t_i$}\label{fig:2}
\end{center}
\end{figure}

Graph of this solution is presented on Fig.~\ref{fig:2}.

\section{Conclusions}\label{6}

In this article, we investigated the non-linear Black--Scholes equation (\ref{FinMath}) from the group theoretic point of view.

First, in Section \ref{2}, using the point transformations of variables (\ref{trans1})--(\ref{trans2}), we reduced the equation to the more simple and canonical form (\ref{ZP}). We found that for this equation there are known several exact solutions. Our main purpose was to carry out the symmetry analysis of the equation in order to obtain a comprehensive list of self-similar (invariant) exact solutions of the one using the method of symmetry reduction to the ordinary differential equations.

In Section \ref{3}, we found the MAI of equation (\ref{ZP}). This algebra is the five-dimensional Lie one, which can be written as a semi-direct sum of a one-dimensional algebra and a four-dimensional solvable ideal. Taking into account the widely known classification of sub algebras of low dimensional Lie algebras \cite{PW} and using the Patera--Winternitz--Zassenhaus algorithm, we found the optimal system of one-dimensional sub-algebras of MAI of equation (\ref{ZP}). Using the ones, which satisfy the necessary conditions of existence of the non-degenerate invariant solutions, we carried out the symmetry reduction of the equation to the ordinary differential equations of the first and second order (see Table \ref{tab:1}) and found several general solutions of the ones. For a number of the reduced equations we could not find the general solutions in the explicit form in elementary functions (see Cases 11--14 in Table \ref{tab:2}).

Using the obtained general solutions of the reduced equations, in Section~\ref{4}, we constructed a set of exact solutions of the Black--Scholes equation under study. The complete list of the solutions is presented in Tables \ref{tab:3} and \ref{tab:4}. Finally, we compared our solutions with the ones found previously.

In Section~\ref{5}, we applied results found in the previous section for solving several BVPs with the governing Black--Scholes equation~\eqref{BSE} in the case $c > 0$. We utilized solution 5 of Table~4 to find exact classical solutions of both the stationary \eqref{5-13}--\eqref{5-15} and non-stationary \eqref{5-24}--\eqref{5-26} BVPs of the Dirichlet and European Call option types, respectively.  

\section*{Funding}

This research did not receive any specific grant from funding agencies in the public, commercial, or not-for-profit sectors.

\end{document}